\global\def\draftcontrol{0}

   \def\versionno{  } 

\catcode`\@=11 

\expandafter\ifx\csname draftcontrol\endcsname\relax\global\def\draftcontrol{0} 
\fi 

{\count255=\time\divide\count255 by 60 
\xdef\hourmin{\number\count255} 
\multiply\count255 by-60\advance\count255 by\time 
\xdef\hourmin{\hourmin:\ifnum\count255<10 0\fi\the\count255}} 
\def\draftdate{\number\month/\number\day/\number\year\ \ \ \hourmin } 

\newcommand\makepapertitle{\par

  \begingroup 
    \renewcommand\thefootnote{\@fnsymbol\c@footnote}%
    \def\@makefnmark{\rlap{\@textsuperscript{\normalfont\@thefnmark}}}%
    \long\def\@makefntext##1{\parindent 1em\noindent 
            \hb@xt@1.8em{%
                \hss\@textsuperscript{\normalfont\@thefnmark}}##1}%
     \newpage 
     \global\@topnum\z@   
     \@makepapertitle 
     \thispagestyle{empty}\@thanks 
  \endgroup 
  \setcounter{footnote}{0}%
  \global\let\thanks\relax 
  \global\let\makepapertitle\relax 
  \global\let\@makepapertitle\relax 
  \global\let\@thanks\@empty 
  \global\let\@author\@empty 
  \global\let\@date\@empty 
  \global\let\@title\@empty 
  \global\let\title\relax 
  \global\let\author\relax 
  \global\let\date\relax 
  \global\let\and\relax 
  \def\version{\let\version\@version\@gobble} 
} 
\def\@makepapertitle{%
  \newpage 
   \ifnum\draftcontrol=1 {} 
   \version\versionno 
   \vskip 5em%
   \else 
   \hfill\hbox to 3cm {\parbox{4cm}{\@pubnum}\hss}%
   \vskip 5em%
   \fi 
   \begin{center}%
   \let \footnote \thanks 
      {\hskip -0\textwidth \hbox to 1\textwidth%
        {\centerline{\Large\bf{\noindent\@title}}}}%
     \vskip 2em%
     {\normalsize
       \lineskip .5em%
       \begin{tabular}[t]{c}%
         \@author 
       \end{tabular}\par}%
     \vskip 1em%
     {\@bstract}%
     \end{center}%
     \vfill
     \@date%
     \vskip 1.5em%
   \par 
} 

\gdef\@pubnum{} 
\def\pubnum#1{%
  \gdef\@pubnum{#1}} 

\gdef\@bstract{} 
\def\Abstract#1{%
  \gdef\@bstract{%
   \parbox{\textwidth-0pc}{%
   \centerline{\bf Abstract}\penalty1000 
   \noindent
   \renewcommand\baselinestretch{1.0} 
   {#1}}} 
} 

\gdef\@email{}
\def\email#1{%
   \gdef\@email{%
   Email: {\tt #1}}
}

\def\ps@paper{\let\@mkboth\@gobbletwo%
     \ifnum\draftcontrol=1 
        \def\@oddfoot{\hbox to \textwidth{\tiny \versionno \hfil\tiny\draftdate}%
        \hskip -\textwidth \hbox to \textwidth{\hfil\rm\thepage\hfil}}%
     \else\def\@oddfoot{\hbox to \textwidth{\hfil\rm\thepage\hfil}} 
     \fi 
     \let\@evenfoot\@oddfoot 
} 

\def\body{\clearpage 
          \pagestyle{paper} 
        } 
\newenvironment{acknowledgments}{%
\vskip 3.25ex 
\noindent {\bf Acknowledgments} 
} 

\def\@version#1{\ifnum\draftcontrol=1 
\typeout{}\typeout{#1}\typeout{} 
\vskip3mm\centerline{\hbox{\fbox{\normalsize{\tt DRAFT -- #1 -- } 
                   {\draftdate}}}}\vskip3mm 
\fi} 
\let\version\@version 
\long\def\eqlabel#1{\ifnum\draftcontrol=1 
                    \tag@false  
                    \tag*{(\theequation) \hbox to -0.2cm{\hspace{0cm}\small{#1}\hss}} 
                    \refstepcounter{equation}  
                    \edef\@currentlabel{\theequation} 
                    \ltx@label{#1}          
                    \else 
                    \label{#1} 
                    \fi 
                    } 
\let\st@bibitem\@bibitem 
\let\st@lbibitem\@lbibitem 
\ifnum\draftcontrol=1 
  \def\@bibitem#1{%
    \st@bibitem{#1}\a@@label{#1}\ignorespaces} 
  \def\@lbibitem[#1]#2{%
    \st@lbibitem[#1]{#2}\a@@label{#2}\ignorespaces} 
  \def\a@@label#1{%
    \gdef\a@lab{\smash{\normalfont\small#1}} 
    \ifvmode 
      \if@inlabel 
        \global\setbox\@labels\hbox{%
          \llap{\a@lab\let\a@lab\relax 
                \kern\@totalleftmargin\kern\marginparsep}%
          \box\@labels}%
      \fi 
    \fi} 
\fi 

\documentclass[12pt,letterpaper]{article} 

\usepackage{amsmath,bm,amsfonts,amssymb,array,calc,amsthm,rotating}
\usepackage[nosort]{cite} 
\usepackage{graphicx}

\renewcommand\baselinestretch{1.25} 
\renewcommand\section{\@startsection {section}{1}{\z@}%
                                   {-3.5ex \@plus -1ex \@minus -.2ex}%
                                   {2.3ex \@plus.2ex}%
                                   {\normalfont\large\bfseries}} 
\renewcommand\subsection{\@startsection{subsection}{2}{\z@}%
                                   {-3.25ex\@plus -1ex \@minus -.2ex}%
                                   {1.5ex \@plus .2ex}%
                                   {\normalfont\normalsize\bfseries}} 
\renewcommand\subsubsection{\@startsection{subsubsection}{3}{\z@}%
                                   {-3.25ex\@plus -1ex \@minus -.2ex}%
                                   {1.5ex \@plus .2ex}%
                                   {\normalfont\normalsize\it}} 
\renewcommand\paragraph{\@startsection{paragraph}{4}{\z@}%
                                   {-3.25ex\@plus -1ex \@minus -.2ex}%
                                   {1.5ex \@plus .2ex}%
                                   {\normalfont\normalsize\bf}} 
\renewcommand\subparagraph{\@startsection{subparagraph}{5}{\z@}%
                                   {-1.25ex\@plus -1ex \@minus -.2ex}%
                                   {0ex \@plus .2ex}%
                                   {\normalfont\normalsize\it}} 

\long\def\@makecaption#1#2{%
  \vskip\abovecaptionskip
  \sbox\@tempboxa{{\bf #1:} #2}%
  \ifdim \wd\@tempboxa >\hsize
    {\small\bf #1:} {\small #2}\par
  \else
    \global \@minipagefalse
    \hb@xt@\hsize{\hfil\box\@tempboxa\hfil}%
  \fi
  \vskip\belowcaptionskip}

\def\ie{{\it i.e.}} 
\def\eg{{\it e.g.}} 

\def\revise#1       {\raisebox{-0em}{\rule{3pt}{1em}}%
                     \marginpar{\raisebox{.5em}{\vrule width3pt\ 
                     \vrule width0pt height 0pt depth0.5em 
                     \hbox to 0cm{\hspace{0cm}{%
                     \parbox[t]{4em}{\raggedright\footnotesize{#1}}}\hss}}}}

\def\calm         {{\cal M}} 
\def\caln         {{\cal N}} 
\def\calo         {{\cal O}}

\def\calr         {{\cal R}}

\def\calw         {{\cal W}} 

\def\complex      {{\mathbb C}} 
 
\def\projective   {{\mathbb P}} 
 
\def\reals        {{\mathbb R}} 
\def\zet          {{\mathbb Z}}

\def\ee           {{\it e}} 
\def\ii           {{\it i}}

\def\Im           {{\rm Im\hskip0.1em}}

\def\sqr#1#2{{\vcenter{\vbox{\hrule height.#2pt   
 \hbox{\vrule width.#2pt height#1pt \kern#1pt 
 \vrule width.#2pt}\hrule height.#2pt}}}}

\def\Hom{\mathop{\rm Hom}}
\def\U{{\it U}}
\def\Ker{\mathop{\rm Ker}}
\def\Im{\mathop{\rm Im}}

\catcode`\@=12 

\begin{document} 


\title{F-term equations near Gepner points}

\pubnum{%
NSF-KITP-04-50 \\ 
hep-th/0404196} 
\date{April 2004} 

\author{
Kentaro Hori$^{a}$ and Johannes Walcher$^{b}$\\[.4cm]
\it ${}^{a}$University of Toronto, Toronto, Ontario, Canada\\[0.2cm]
\it ${}^b$Kavli Institute for Theoretical Physics \\
\it University of California, Santa Barbara, California, USA \\[0.4cm] 
}


\Abstract{We study marginal deformations of B-type D-branes in 
Landau-Ginzburg orbifolds. The general setup of matrix factorizations 
allows for exact computations of F-term equations in the low-energy 
effective theory which are much simpler than in a
corresponding geometric description. We present a number of obstructed
and unobstructed examples in detail, including one in which 
a closed string modulus is obstructed by the presence of D-branes.
In a certain example, we find a non-trivial global structure of
the BRST operator on the moduli space of branes.
}


\makepapertitle 

\body 

\version\versionno 


\section{Introduction}

D-branes in non-trivial Calabi-Yau(CY) backgrounds
are interesting to study and 
find many applications throughout string theory. The subject has been
investigated intensively over the past half decade, with many remarkable
results. One part of the story that is still 
less understood are the properties of D-brane moduli spaces, both at
the local and in particular at the global level. Some progress has 
been obtained in the context of non-compact toric models, but the 
computation of D-brane superpotentials for a generic compact CY model
is in general still missing, despite considerable interest in 
particular from the phenomenological point of view.

A large class of string compactification backgrounds admit in some 
part of closed string moduli space a description as Landau-Ginzburg 
orbifold (LG) models. In the bulk, such LG models are specified by 
the choice of a quasihomogeneous polynomial $W$ as worldsheet 
superpotential. Since \cite{warner}, finding 
useful B-type boundary conditions in Landau-Ginzburg models was 
a nagging problem (see for example \cite{gjs,hiv,linear,hklm}).
Recently,
M. Kontsevich has given a description based on 
a so-called matrix factorization of $W$ \cite{Kontsevich}
which is found to be useful in more recent works
\cite{kali1,orlov,bhls,kali2,kali3,calin27,aad,hori,hll,hl}.
It can be succinctly written as the equation
\begin{equation}
Q^2 = W\cdot {\rm id} \,,
\eqlabel{master}
\end{equation}
on a matrix $Q$ with polynomial entries which encodes boundary 
interactions on the worldsheet.

The purpose of this note is to open up this window on the possibility 
of studying D-brane moduli spaces and D-brane superpotentials in this 
very simple algebraic setup. We will firstly discuss some general
aspects of the deformation problem of D-branes in their description
as matrix factorizations. We focus on marginal deformations which are
relevant for the moduli problem. (Relevant deformations,
important for discussion on (in)stability and decays,
were studied in \cite{hori,hll}.)
Secondly, we will apply the technology to 
a number of relevant examples. We will start with the case $\hat c=1$, 
corresponding geometrically to an elliptic curve. We will show how
the moduli space of its matrix factorizations is naturally the torus
itself. Results from the mathematical literature \cite{laza}
suggest that the
problem of finding all matrix factorizations for the torus case is
essentially solved.
In this example, we
also find that the matrix $Q$, which is a part of the
BRST operator in the context
of string field theory, transforms non-trivially as we move
around in the moduli space.
We then turn to the physically interesting case
$\hat c=3$. We study in detail the quintic model and the behavior
of its rational branes under open and closed string deformations.
One spinoff of our results is the reconciliation, via mirror symmetry,
of the behavior of A-branes wrapped on $\reals\projective^3$ inside 
the quintic over K\"ahler moduli space.
We will also find an example where a
certain deformation of closed string moduli is obstructed by
the presence of a D-brane.
Such a phenomenon is known to be possible on general grounds 
but a concrete example had not been found
in the context of 4d $\caln=1$ supersymmetry.

{\it Note:} This publication was prompted by the preprint \cite{rutgers}, 
which appeared while we were contemplating publication of our results,
and in which the idea of using B-type LG branes for computing 
spacetime superpotentials is also discussed.

\section{Generalities}
\label{general}

We consider two-dimensional $\caln=2$ Landau-Ginzburg theories of relevance
for superstring compactifications (LG models). We propose to study conformally 
invariant B-type boundary conditions as worldsheet descriptions of D-branes 
in such models.

\subsection{The bulk}

The construction of such LG models begins with picking a superpotential
$W$, which is a holomorphic function of $r$ chiral field variables $x_i$.
Since we require a conformally invariant IR fixed point, we
take $W$ to be a quasihomogeneous polynomial
\cite{vafa1,martinec,vawa,WittenLG}, of 
degree denoted by $H$, where each $x_i$ has weight $w_i$, \ie,
\begin{equation}
W(\lambda^{w_1} x_1,\ldots, \lambda^{w_r} x_r) = 
\lambda^H W(x_1,\ldots,x_r) \qquad \text{for $\lambda\in\complex$}
\eqlabel{qh}
\end{equation}
The central charge of the theory at the IR fixed point is given by 
\begin{equation}
\hat c = \sum_{i} \bigl(1- \frac{2 w_i}H\bigr) = \sum_i \bigl(1-\frac 2{h_i}\bigr)
\eqlabel{central}
\end{equation}
We will mostly assume that all $w_i$ divide $H$, so that $h_i$ are integer, and
there exists a Fermat point in the moduli space of conformal theories, at which 
$W$ takes the form,
\begin{equation}
W = x_1^{h_1}+\cdots+x_r^{h_r}\,.
\eqlabel{fermat}
\end{equation}
At this Fermat point, the IR fixed point theory is rational and equivalent to 
a tensor product of $\caln=2$ minimal models, generically refered to as a Gepner 
model \cite{gepner}.

The chiral $(c,c)$ ring \cite{chiral} of such an LG theory is given by
\begin{equation}
\calr = \complex[x_i] / \langle dW\rangle
\eqlabel{ccring}
\end{equation}
and because of \eqref{qh} is graded by the vector $\U(1)$ R-charge $q$, 
which is normalized such that $W$ has charge $2$,
\begin{equation}
q_i=q(x_i) = \frac 2{h_i}
\end{equation}

Let us also assume that the central charge \eqref{central}, which measures the
number of compactified dimensions, is integer. As is well-known (see, \eg,
\cite{vafa2}) we can then project onto integral $\U(1)$ charge by orbifolding 
by the global symmetry group $\Gamma_0\cong\zet_H$ whose generator acts by
\begin{equation}
x_i\mapsto \omega^{w_i} x_i
\end{equation}
with $\omega^H=1$. In the RCFT at the Fermat point, this orbifold operation is
most conveniently phrased in terms of simple-current extensions, but we
will not use this language here. 

Orbifolding thus projects the $(c,c)$ ring \eqref{ccring} onto
integrally charged
fields. Twisted sectors contain the $(a,c)$ ring as well as possibly additional
elements of the $(c,c)$ ring.
We often enlarge our orbifold group $\Gamma$
to contain other global symmetries as well.
The maximal example is the one $\widehat{\Gamma}_0$ for
Greene-Plesser mirror \cite{grpl} of $\Gamma_0$, $x_i\to \omega_ix_i$
where $\omega_1^{h_1}=\cdots=\omega_r^{h_r}=\omega_1\cdots\omega_r=1$.
More generally, $\Gamma$ is included in $\widehat{\Gamma}_0$ and includes
$\Gamma_0$.

\subsection{B-type Boundary conditions}
\label{Btype}

As explained in
\cite{kali1,bhls,kali2,calin27} the 
problem of finding boundary conditions preserving B-type $\caln=2$ 
supersymmetry in an LG model becomes equivalent to finding a matrix $Q=Q(x_i)$ 
satisfying \eqref{master} \cite{Kontsevich}.
To explain \eqref{master} more precisely, we 
are looking for a square matrix $Q$ whose entries are polynomials in $x_i$, 
such that there exists a grading operator $\sigma$ (a matrix with scalar 
entries, $\sigma^2=1$) satisfying
\begin{equation}
\sigma Q + Q \sigma =0 \,.
\end{equation}
$Q$ can be thought of as acting on a $\zet_2$ graded $\complex[x_i]$-module
(with grading provided by $\sigma$), which we will denote by $N$. The RHS 
of equation \eqref{master} is interpreted as the bulk LG potential $W$ 
times the identity matrix. Equivalently, by diagonalizing $\sigma$, we 
are looking for a pair of matrices $f$ and $g$ such that
\begin{equation}
Q = \begin{pmatrix} 0 & f \\ g & 0\end{pmatrix} \qquad 
\left[ \sigma=\begin{pmatrix} 1 & 0 \\ 0 & -1 \end{pmatrix}\right] \,,
\end{equation}
in terms of which \eqref{master} becomes
\begin{equation}
fg = g f = W \cdot {\rm id} \,.
\eqlabel{factorize}
\end{equation}

The physical interpretation of this formalism is that $Q$ represents a tachyon
configuration between a stack of spacefilling branes and a stack of spacefilling
antibranes, corresponding to the positive and negative eigenspaces of $\sigma$,
respectively.

Because of the form \eqref{factorize}, a solution of \eqref{master} is also
known as a matrix factorization in the mathematical literature. We will 
use this terminology as well as referring to the triple $(N,\sigma,Q)$ as a
B-type LG brane, often dropping $Q$ and $\sigma$.

Generally, we do not definitely want to fix the dimension of $Q$ (which is
even and twice the dimensions of $f$ and $g$). This is because we actually
want to identify solutions which differ by the addition of a "trivial 
brane-antibrane pair" corresponding to
\begin{equation}
Q_{\rm trivial} = \begin{pmatrix} 0 & 1 \\ W & 0\end{pmatrix}
\end{equation}
In other words we want to divide the space of solutions of \eqref{master}
by the equivalence relation
\begin{equation}
Q \equiv Q\oplus Q_{\rm trivial}
\eqlabel{trivial}
\end{equation}

We also want to identify, of course, solutions of \eqref{master} which
differ only by a similarity transformation
\begin{equation}
Q \mapsto U Q U^{-1}
\eqlabel{gauge}
\end{equation}
where $U$ is an invertible matrix with polynomial entries.

As in the bulk, one expects that the boundary interactions defined by
such a matrix factorization will flow to a conformally invariant boundary
field theory in the IR if there exists a conserved $\U(1)$ R-charge.
Since the $\U(1)$ charge of the $x_i$ is fixed from the bulk, the only 
freedom we have is the action on the CP spaces. Thus, we require the
existence of a matrix $S$ such that $Q$ has definite charge under it,
\begin{equation}
\ee^{\ii \alpha S} Q\bigl(\ee^{\ii \alpha q_i}x_i\bigr) 
\ee^{-\ii\alpha S} = \ee^{\ii\alpha q(Q)} Q(x_i)
\eqlabel{rcharge}
\end{equation}
It is clear from \eqref{master} that if we normalize $W$ to have charge $2$,
$q(Q)=1$. The condition \eqref{rcharge} is the boundary equivalent of
\eqref{qh}, and $S$ is part of the data specifying a conformally invariant
B-type boundary condition.

The spaces of boundary chiral fields from one brane $(M,P,\rho)$ to
another such brane $(N,Q,\sigma)$ (for the same bulk superpontential $W$)
is obtained from the space
\begin{equation}
{\Hom}_{\complex[x_i]}(M,N) \,,
\end{equation}
by taking the cohomology of the operator $D$ defined by
\begin{equation} 
D (\Phi) = Q\Phi + \sigma \Phi \rho P\,.
\end{equation}
for an arbitrary $\Phi=\Phi(x_i)$ (a matrix with polynomial entries).
We will denote the space by
\begin{equation}
H^*(M,N) = \Ker (D)/\Im(D)
\end{equation}

In general, this space is $\zet_2$ graded by $\sigma$ and $\rho$, \ie,
homogeneous elements satisfy
\begin{equation}
 \sigma\Phi\rho = (-1)^{\Phi} \Phi
\end{equation}
However, in the conformally invariant case, we have another, finer, (in general
non-integral!) grading by $\U(1)$ R-charge. Denoting the R-charges associated 
with $(M,P,\rho)$ and $(N,Q,\sigma)$ by $R$ and $S$ respectively, we can 
contemplate homogeneous elements satisfying
\begin{equation}
\ee^{\ii\alpha S}\,\Phi\bigl(\ee^{\ii\alpha q_i} x_i\bigr)\, \ee^{-\ii\alpha R} = 
\ee^{\ii \alpha q(\Phi)}\Phi(x_i) \,.
\eqlabel{rfield}
\end{equation}
We also note that imposing unitarity of the worldsheet theory requires
$0\le q\le \hat c$ for all chiral fields \cite{chiral}. Clearly, this is
a condition on the matrices $R$ and $S$ in addition to \eqref{rcharge}.
Similarly to the bulk, charge conjugation invariance of the open string RR sector 
translates into "Serre duality" for boundary chiral fields. We will denote it 
by $\dagger$. It maps $\Phi\in H^*(M,N)$ to $\Phi^\dagger\in H^*(N,M)$
and satisfies
\begin{equation}
q(\Phi^\dagger) = \hat c - q(\Phi)
\eqlabel{serre}
\end{equation}

Looking back at the bulk, we see that the next step in the construction is
orbifolding. On the boundary, this is again implemented by the choice of
an action on the CP spaces. Choosing a representation of $\Gamma$ for 
each brane $(M,\sigma,Q)$, we impose equivariance on the factorization $Q$
\begin{equation}
\gamma \, Q\bigl(\gamma(x_i)\bigr) \gamma^{-1} = Q(x_i) \qquad\text{for every 
$\gamma\in\Gamma$}
\end{equation}
Open string fields are projected similarly. It must be that after such
a projection all fields from a brane to itself have integer R-charge, which upon
$\bmod 2$ reduction is the same as the $\zet_2$ grading.

\subsection{Deformations and Obstructions}
\label{deform}

Let us now fix one such B-brane $(M,\sigma,Q)$ in some appropriately 
orbifolded bulk theory with bulk superpotential $W$. We want to ask the 
following questions

(i) Can we deform $Q$, holding $W$ fixed?

(ii) If we deform $W$, is there a deformation of $Q$ that satisfies 
\eqref{master}?

\noindent
If we restrict to infinitesimal deformations, the answer to question (i)
is simple. Infinitesimal deformations correspond to elements
\begin{equation}
\Phi \in H^1(M,M) \,,
\end{equation}
since clearly the deformed $Q(\varphi)=Q+\varphi\Phi$ satisfies
\begin{equation}
\bigl(Q(\varphi)\bigr)^2 = Q^2 + \varphi\{Q,\Phi\} + \varphi^2 \Phi^2 
= W + \calo(\varphi^2) \,,
\eqlabel{first}
\end{equation}
\ie, it squares to $W$ to first order in $\varphi$. It is easy to see that 
we cannot remove $\varphi\Phi$ from $Q(\varphi)$ by a gauge transformation,
unless $\Phi$ is trivial in $H^1(M,M)$.

Let us try to continue the first order deformation \eqref{first} to
higher order in $\varphi$, \ie, we write
\begin{equation}
Q(\varphi) = \sum_{n} \varphi^n Q_n \,,
\eqlabel{higher}
\end{equation}
where $Q_0=Q$ we started with, $Q_1=\Phi$, and all $Q_n$ have odd degree
and R-charge $1$. Imposing $\bigl(Q(\varphi)\bigr)^2=W$ then leads at order 
$n$ to the equation
\begin{equation}
\{Q_0,Q_n\} = - \sum_{k=1}^{n-1} Q_k Q_{n-k} \,.
\eqlabel{ordern}
\end{equation}
Assume that we have found a deformation up to order $n-1$. The RHS of 
\eqref{ordern} is then $Q_0$ closed,
\begin{equation}
\bigl\{Q_0,\textstyle{\sum_{k=1}^{n-1} Q_k Q_{n-k}}\bigr\} =
\displaystyle
- \sum_{k=1}^{n-1} \biggl( \sum_{l=1}^{k-1} Q_l Q_{k-l} Q_{n-k}
- \sum_{l=1}^{n-k-1} Q_k Q_l Q_{n-k-l}\biggr) = 0 \,.
\end{equation}
Thus, we can solve to order $n$ unless the RHS of \eqref{ordern}
is in the cohomology of $Q_0=Q$. Our problem being $\zet$-graded,
the RHS has degree $2$ so the possible obstructions lie in
$H^2(M,M)$. 

A simple consequence of these considerations is the dependence of the
deformation problem on the dimension, or central charge $\hat c$ of
our model. For $\hat c=1$ (compactification on a torus), $H^p$ 
vanishes for $p>1$, so the deformation problem is never obstructed. 
We will see explicitly in a later section how this is implemented in
practice.

The case $\hat c=3$ is the most interesting one from the physics
point of view. We note that in that case, Serre duality \eqref{serre} 
implies that to every first order deformation $\Phi\in H^1(Q)$, there 
exists a corresponding obstruction $\Phi^\dagger\in H^2(Q)$.
``Generically'', one would expect that all obstructions appear in the
deformation problem. As we will see, however, this does not mean
that there are no finite boundary deformations for $\hat c=3$.
Instead, it can happen that the boundary obstructions actually serve
to lift a previously marginal bulk deformation! 

Thus turning to problem (ii), we consider a bulk deformation
$W\to W+\psi\Psi$, where $\Psi$ is a polyonmial in $x_i$ of total degree 
$H$, \ie, left-right $\U(1)$-charge $q_L=q_R=1$. Multiplying $\Psi$ with
the identity matrix transforms it into a boundary field with R-charge
$q=q_L+q_R=2$. Obviously, $\Psi\cdot {\rm id}$ is $Q$-closed, so with 
the ansatz
\begin{equation}
Q(\psi) = \sum \psi^n Q_n
\eqlabel{ansatz}
\end{equation}
and $Q_0=Q$, we can solve \eqref{master} to first order in $\psi$ only 
if 
\begin{equation}
\{Q_0,Q_1\} = \Psi\,{\rm id}
\end{equation}
\ie, $[\Psi]=0\in H^2(Q)$. In this case, we obtain at order $n$ in $\psi$
the condition
\begin{equation}
\{Q_0,Q_n\} = - \sum_{k=1}^{n-1} Q_k Q_{n-k}
\eqlabel{ordernn}
\end{equation}
Superficially, this looks like \eqref{ordern}. And indeed, even though
$Q_1$ is not closed, the argument we gave above still goes through (because
$\Psi$ commutes with everybody on the boundary) to show that if we have
solved to order $n-1$, we can solve to order $n$ if
and only if the RHS of \eqref{ordernn}
is trivial in $H^2(Q)$.

On the other hand, if $\Psi$ is non-trivial in $H^2(Q)$,
we will not be able to 
solve \eqref{master}, at least not with the ansatz \eqref{ansatz}. We will
see in an example that what can happen in that case is that there is a 
first order boundary deformation $\Phi$ which squares to $\Psi$. In that 
case we obtain two families of solutions
\begin{equation}
\bigl(Q + \varphi\Phi\bigr)^2 = W + \psi \Psi
\end{equation}
for $\varphi^2 = \psi$. More generally, $\Psi$ might appear as an
obstruction to a boundary deformation $\varphi\Phi$ at some higher order
$n>2$. In this case, we obtain a polynomial constraint between $\psi$
and $\varphi$ of order $n$.

Still, it is possible that in the general case in which a bulk
field $\Psi$ restricts on the boundary to an element of $H^2(Q)$, there
is no corresponding boundary deformation that yields a solution of
\eqref{master}.
In that case, that bulk deformation is obstructed by the presence of
the D-brane. We will find such an example in Section~\ref{Obulk}.

On the other hand, it should also be noted that not all obstructions on the
boundary can be lifted by deforming the bulk. This is because, quite 
simply, not all boundary fields of charge $2$ are proportional to the 
identity matrix and can be moved to the bulk.

\subsection{F-terms and Superpotentials}

$\caln=2$ superconformal symmetry and charge intgrality
on the worldsheet is the condition for
spacetime $\caln=1$ supersymmetry 
at string tree level, described by
the D-term and F-term equations in the low energy effective supergravity.
We have discussed in the previous subsection that studying the deformation
problem for B-branes naturally leads to some holomorphic constraints on 
the open and closed deformation parameters $\varphi$ and $\psi$.
These deformation parameters become $\caln=1$ chiral fields 
in the low-energy theory.

The constraints on the
fields $\varphi$ and $\psi$ are then naturally interpreted as F-term 
equations. On general grounds, see \eg,
\cite{BDLR}, F-terms are related to $\caln=2$ supersymmetry on the
worldsheet, and since \eqref{master} is equivalent to preserving
$\caln=2$ worldsheet supersymmetry, we conclude that these constraints
on $\varphi$ and $\psi$ we find from studying brane deformations are 
all F-term constraints in the tree level low-energy theory.
On the other hand, the requirement of conformal invariance, or,
equivalently, conditions \eqref{rcharge} and \eqref{rfield} on the 
R-charge, are related to D-terms in spacetime \cite{BDLR,kklm1}.

Given the F-term constraints, it is natural to ask whether one
can integrate them to obtain an $\caln=1$ spacetime superpotential
$\calw(\varphi,\psi)$ governing the dynamics of light open and closed 
string fields. 
 One requirement is certainly that finding the locus of
$\calw=d\calw=0$ must correspond to solving \eqref{master}. However,
it is also important that $\calw$ be expressed in the natural
``flat variables'', \eg, in order for it to be useful for mirror 
symmetry. Moreover, there are certain global requirements on 
$\calw$ that must be taken into account, such as that it contain 
all fields that can become massless at some point in the moduli 
space, as well as that it take value in the right bundle over 
configuration space. We will mention some of these global 
conditions in the examples below.

General prescriptions for the computation of $\calw$ 
from the topological string theory
 have been discussed, \eg, in \cite{calin1,toma,dgjt,hll}
(the underlying mathematical literature is \cite{FOOO}).
The results of \cite{vafa3,kali2} on
topological correlators in LG model also
 should be useful.

\section{Constructions}

\subsection{Minimal models}

B-branes in the $\caln=2$ minimal models have been studied from the
point of view that we take here in particular in \cite{bhls,kali3,
hori,hll}.

In the minimal model with $W=x^h$, we denote the B-brane associated
with the factorization $W=x^n x^{h-n}$ by $(M_n,Q_n)$, where 
$M_n$ is a rank two free module and
\begin{equation}
Q_n=Q_n(x) = \begin{pmatrix}
0 & x^n\\ x^{h-n} & 0
\end{pmatrix}.
\end{equation}
The fermionic and bosonic
operators between $M_{n_1}$ and $M_{n_2}$ are given
by
\begin{equation}
\phi^1_{n_1,n_2,j}(x) =
\begin{pmatrix}
0 & x^{\frac{n_1+n_2}{2}-j-1} \\ -x^{h-\frac{n_1+n_2}{2}-j-1} & 0
\end{pmatrix}\,,
\end{equation}
and
\begin{equation}
\phi^0_{n_1,n_2,j}(x) =
\begin{pmatrix}
x^{j-\frac{n_1-n_2}{2}} & 0\\ 0 & x^{j+\frac{n_1-n_2}{2}} 
\end{pmatrix}\,,
\end{equation}
where
$$
j=\frac{|n_1-n_2|}{2},\frac{|n_1-n_2|}{2}+1,\ldots,
\min\left\{\frac{n_1+n_2}{2}-1,h-\frac{n_1+n_2}{2}-1\right\}.
$$
The R-charge matrix is given by
$$
R_n=\left(\begin{array}{cc}
\frac{1}{2}-\frac{n}{h}&0\\
0&-\frac{1}{2}+\frac{n}{h}
\end{array}\right).
$$
This is determined by the invariance of the boundary interaction
\cite{hori}, up to a shift by matrix proportional to the identity.
This choice is such that the Serre duality holds:
The fermionic field $\phi^1_{n_1,n_2,j}(x)$ and
the bosonic field $\phi^0_{n_1,n_2,j}(x)$ have
R-charges $1-\frac{2+2j}{h}$ and $\frac{2j}{h}$ respectively,
and $\phi^1_{n_1,n_2,j}(x)$ and $\phi^0_{n_2,n_1,j}(x)$
are indeed Serre dual of each other.

The $\zet_h$ symmetry $x\to \omega x$, $\omega^h=1$, induces actions on
the Chan-Paton factor $M_n$.
They are labeled by a mod $2h$ integer
$m$ such that $n+m$ is even;
\begin{equation}
\gamma_{n,m}(\omega)=
\begin{pmatrix}
\omega^{-\frac{n+m}{2}}&0\\
0&\omega^{\frac{n-m}{2}}
\end{pmatrix}.
\end{equation}
We denote by $M_{n,m}$ the B-brane $M_n$ equipped with
this $\zet_h$-action. The $\zet_h$ symmetry acts on
the fields between $M_{n_1,m_1}$ and $M_{n_2,m_2}$
by $\phi(x)\mapsto \gamma_{n_2,m_2}(\omega)\phi(\omega x)
\gamma_{n_1,m_1}(\omega)^{-1}$.
In particular, the chiral fields are transformed as follows
\begin{equation}
\phi^0_{n_1,n_2,j}(x)\mapsto \omega^{j+\frac{m_1-m_2}{2}}
\phi^0_{n_1,n_2,j}(x),\qquad
\phi^1_{n_1,n_2,j}(x)\mapsto \omega^{-j-1+\frac{m_1-m_2}{2}}
\phi^1_{n_1,n_2,j}(x).
\label{orbac}
\end{equation}

\subsection{Tensor products}

As the first step toward Gepner model,
we construct the tensor products of the above elementary 
factorizations. By ``tensor product'', we here mean in the graded
sense. 
Slightly formally, this graded tensor product differs from
the ordinary tensor product only in that composition of maps
respects the grading. If $\otimes$ denotes the ordinary tensor product
and $\odot$ the graded version, then for graded vector spaces
$(M_1,\rho_1)$ and $(M_2,\rho_2)$, we have simply $(M_1,\rho_1)\odot 
(M_2,\rho_2)\cong (M_1\otimes M_2,\rho_1\otimes\rho_2)$. However
for morphisms $\phi_i:(M_i,\rho_i)\longrightarrow (N_i,\sigma_i)$, we 
have
\begin{equation}
\phi_1\odot \phi_2 = \phi_1 \rho_1^{\phi_2} \otimes \phi_2 \,,
\end{equation}
such that composition satisfies 
\begin{equation}
(\phi_1\odot\phi_2) (\psi_1\odot\psi_2) =
(-1)^{\phi_2\psi_1}\phi_1\psi_1\odot\phi_2\psi_2
\eqlabel{compo}
\end{equation}
(for homogeneous maps $\phi_2$ and $\psi_1$).

Explicitly, given a matrix factorization $(N_1,\sigma_1,Q_1)$ of 
$W_1$ and $(N_2,\sigma_2,Q_2)$ of $W_2$, we have the graded
tensor product
\begin{equation}
\begin{split}
(N_1,\sigma_1,Q_1)\odot (N_2,\sigma_2,Q_2) &=
(N_1\odot N_2,\sigma_1\odot \sigma_2,Q_1\odot 1+1\odot Q_2) \\
&=
(N_1\otimes N_2,\sigma_1\otimes \sigma_2,Q_1\otimes 1 + \sigma_1\otimes Q_2)\,.
\end{split}
\end{equation}
It is trivial to check that $Q=Q_1\otimes 1+\sigma_1\otimes Q_2$
squares to $W=W_1+W_2$.

One can check in general \cite{aad} that for two such tensor products 
$(M,\rho,P)=(M_1,\rho_1,P_1)\odot (M_2,\rho_2,P_2)$
and 
$(N,\sigma,Q)=(N_1,\sigma_1,Q_1)\odot (N_2,\sigma_2,Q_2)$
of $W=W_1+W_2$, we have the K\"unneth formula on the cohomologies,
\begin{equation}
H^*(M_1\odot M_2,N_1\odot N_2)
\cong H^*(M_1,N_1)\odot H^*(M_2,N_2) \,.
\eqlabel{kunneth}
\end{equation}

It is also clear that the general considerations concerning R-charges
and orbifold projection described in subsection \ref{Btype} are
compatible with taking graded tensor products. In particular,
\eqref{kunneth} is graded by R-charge in that context.
In the remainder of the paper, we will revert to denoting the 
tensor product of matrix factorizations by $\otimes$.

\subsection{Orbifolds}

Branes in the orbifold model are labeled by
${\bf n}=(n_1,...,n_r)$ for the tensor product brane
and ${\bf m}=(m_1,...,m_r)$
which specifies the orbifold group action on the Chan-Paton factor.
The chiral fields between two such branes are simply the
fields which are invariant under
the orbifold group action \cite{hori,aad}.

Let us consider the boundary preserving sector,
${\bf n}_1={\bf n}_2={\bf n}$, ${\bf m}_1={\bf m}_2$.
The chiral field
$\otimes_{i=1}^n\phi^{s_i}_{n_i,n_i,j_i}(x_i)$
transforms under $(\omega_1,\ldots,\omega_r)\in \Gamma$
by the phase
\begin{equation}
\prod_{s_i=0}\omega_i^{j_i}\prod_{s_i=1}\omega_i^{-j_i-1}.
\end{equation}
See the action (\ref{orbac}).
Since $\Gamma$ always includes the element
$(\omega_1,...,\omega_r)$
with $\omega_i=e^{\frac{2\pi i}{h_i}}$,
 $\sum_{s_i=0}\frac{j_i}{h_i}-\sum_{s_i=1}\frac{j_i+1}{h_i}$
must be an integer for an invariant field
$\otimes_i\phi^{s_i}_{n_i,n_i,j_i}(x_i)$.
The R-charge of such a field
is
\begin{equation}
q=\sum_{s_i=0}\frac{2j_i}{h_i}
+\sum_{s_i=1}\left(1-\frac{2j_i+2}{h_i}\right)
=\#\{i|s_i=1\}
+2\left(\sum_{s_i=0}\frac{j_i}{h_i}-\sum_{s_i=1}\frac{j_i+1}{h_i}\right),
\end{equation}
which is indeed an even integer (resp.\ odd integer) if
the field is bosonic (resp.\ fermionic).
We also note that the Serre dual is obtained by flipping the $s_i$-label,
\begin{equation}
\otimes_i\phi^{s_i}_{n_i,n_i,j_i}(x_i)
\,\stackrel{\rm Serre}{\longleftrightarrow}\,
\otimes_i\phi^{1-s_i}_{n_i,n_i,j_i}(x_i).
\end{equation}
$\Gamma$-invariance of the two sides are equivalent
since $(\omega_1,...,\omega_r)\in \Gamma$ obeys
$\omega_1\cdots\omega_r=1$.

In what follows, we usually drop the ${\bf m}$-labels since
we mainly consider the sectors with ${\bf m}_1={\bf m}_2$.

\section{The torus}

In this section, we consider matrix factorizations of the LG potential
for the two-dimensional torus,
\begin{equation}
W= x_1^3+x_2^3+x_3^3 +\psi x_1x_2x_3
\eqlabel{torus}
\end{equation}
with modulus $\psi$. All our constructions will be $\zet_3$ equivariant,
but we do not make this explicit. 
As mentioned in Section~\ref{deform},
first order deformations should not be obstructed since there is no
obstruction class in the model with $\hat{c}=1$.
Even in such a case, whether the series
$\sum_n\varphi^nQ_n$ has a finite radius of convergence
is a non-trivial problem \cite{Kodaira}.
We will in fact find finite deformations of rational branes,
both at $\psi=0$ and also for finite $\psi$, and observe
that a non-trivial global geometry of the moduli space
of branes emerges.

Our results depend crucially on a mathematical literature
\cite{laza} in which matrix factorizations of \eqref{torus} 
(for $\psi=0$) have been studied and, as we understand, classified.
This will also provide us to find a clue on geometric interpretation of
the Landau-Ginzburg branes.

\subsection{A family of matrix factorizations}

Here, we consider a particular family of matrix factorizations
which reduces in a limit to the tensor product of minimal model 
branes. This solution is obtained by utilizing results from 
\cite{laza}. Consider the matrix 
\begin{equation}
A=
\begin{pmatrix}
\alpha x_1 & \beta x_3 & \gamma x_2\\
\gamma x_3 & \alpha x_2 & \beta x_1 \\
\beta x_2 & \gamma x_1 & \alpha x_3
\end{pmatrix}
\eqlabel{A}
\end{equation}
We see that
\begin{equation}
\det A = (\alpha^3+\beta^3+\gamma^3) x_1x_2 x_3 - \alpha\beta\gamma
(x_1^3+x_2^3+x_3^3)
\end{equation}
which is equal to $\lambda W$ with
\begin{equation}
\lambda = -\alpha\beta\gamma
\end{equation}
if and only if
\begin{equation}
\alpha^3+\beta^3+\gamma^3 +\psi \alpha\beta\gamma = 0.
\eqlabel{alpha}
\end{equation}
Let $B$ be the adjoint of $A$ up to a factor,
\begin{equation}
\begin{split}
B &:= \frac 1\lambda \mathop{{\rm adj}}(A) \\
&= -\frac 1{\alpha\beta\gamma}
\begin{pmatrix}
\alpha^2 x_2x_3 -\beta\gamma x_1^2 &
\gamma^2 x_1x_2-\alpha\beta x_3^2 &
\beta^2x_1x_3-\alpha\gamma x_2^2 \\
\beta^2x_1x_2-\alpha\gamma x_3^2 &
\alpha^2 x_2x_3-\beta\gamma x_2^2 &
\gamma^2x_2x_3-\alpha\beta x_1^2 \\
\gamma^2 x_1x_3 -\alpha\beta x_2^2 &
\beta^2 x_2x_3-\alpha\gamma x_1^2 &
\alpha^2x_1x_2-\beta\gamma x_3^2
\end{pmatrix}
\end{split}
\end{equation}
Then we find \begin{equation}
A B = B A = W \, {\rm id},
\end{equation}
as long as $(\alpha,\beta,\gamma)$ obeys (\ref{alpha})
and $\alpha\beta\gamma$ is non-zero.
This matrix factorization becomes singular as $\lambda\to 0$,
but we can take the limit by using a trick. Let us consider 
$\alpha\to 0$, $\beta/\gamma\to -1$ as an example. We begin by adding
a trivial brane-antibrane pair
\begin{equation}
f= 
\begin{pmatrix}
-\frac 1\alpha W & 0\\ 0 &A
\end{pmatrix}
\qquad
g=
\begin{pmatrix}
-\alpha & 0 \\ 0 & B
\end{pmatrix}
\end{equation}
and make some elementary transformations such as to remove the singular
part of $B$. The point is that the singular piece of $B$ is of the
form
\begin{equation}
B \sim \frac 1\alpha 
\begin{pmatrix}
x_1^2 & x_1x_2 & x_1x_3\\ x_1x_2 & x_2^2 & x_2 x_3 \\ x_1 x_3 & x_2 x_3 & x_3^2
\end{pmatrix} + {\rm regular}
= \frac1\alpha
\begin{pmatrix}
x_1\\x_2\\x_3
\end{pmatrix}
\cdot
\begin{pmatrix}
x_1&x_2&x_3
\end{pmatrix}
+{\rm regular}
\end{equation}
So we consider 
\begin{equation}
f \to \tilde f=U^{-1 T} f U^{-1}\qquad
g \to \tilde g=U g U^{T}
\end{equation}
where 
\begin{equation}
U=
\begin{pmatrix}
1&0&0&0\\
-x_1/\alpha & 1 & 0&0\\
-x_2/\alpha &0&1&0\\
-x_3/\alpha &0&0&1
\end{pmatrix}
\label{basischange}
\end{equation}
and find
\begin{equation}
\tilde g=
\begin{pmatrix}
-\alpha & x_1&x_2&x_3\\
x_1 & 
-\frac{\alpha}{\beta\gamma} x_2x_3 & 
\frac{x_3^2}\gamma - \frac1\alpha\bigl(1+\frac\gamma\beta\bigr)x_1x_2 &
\frac{x_2^2}\beta - \frac1\alpha\bigl(1+\frac\beta\gamma\bigr)x_1x_3 \\
x_2 & 
\frac{x_3^2}\beta - \frac1\alpha\bigl(1+\frac\beta\gamma\bigr)x_1x_2 &
-\frac{\alpha}{\beta\gamma} x_1x_3 &
\frac{x_1^2}\gamma - \frac1\alpha\bigl(1+\frac\gamma\beta\bigr)x_2x_3  \\
x_3 &
\frac{x_2^2}\gamma - \frac1\alpha\bigl(1+\frac\gamma\beta\bigr)x_1x_3 &
\frac{x_1^2}\beta - \frac1\alpha\bigl(1+\frac\beta\gamma\bigr)x_2x_3 &
-\frac{\alpha}{\beta\gamma} x_1x_2
\end{pmatrix}
\end{equation}
This is nonsingular in the limit $\alpha\to 0$,
$\beta/\gamma\to -1$. In fact, it is easy to see from \eqref{alpha}
\begin{equation}
\beta+\gamma = \frac\psi 3\alpha + \calo(\alpha^3)
\eqlabel{limit}
\end{equation}
For $\psi=0$, the limit indeed reduces to the matrix
factorization obtained from the taking product of minimal models
$M_1(x_1)\otimes M_1(x_2)\otimes M_1(x_3)$ and perturbing by the 
marginal operator $\Phi=\phi^1_0(x_i)^{\otimes i=1\ldots3}$.
Indeed, setting $\beta=1$ for convenience, we find,
\begin{equation}
\tilde g \sim
\begin{pmatrix}
0& x_1&x_2&x_3 \\
x_1 & 0 & -x_3^2 & x_2^2 \\
x_2 & -x_3^2 & 0 & -x_1^2 \\
x_3 & -x_2^2 & x_1^2 & 0 
\end{pmatrix}
+ \alpha
\begin{pmatrix}
-1& 0 & 0 & 0\\
0 & x_2 x_3 & 0 & 0\\
0 & 0 & x_1 x_3 & 0\\
0 & 0 & 0 & x_1 x_2 
\end{pmatrix}
+\calo(\alpha^2)
\end{equation}
Let us also write out the other matrix after the similarity
transformation
\begin{equation}
\tilde f = 
\begin{pmatrix}
\rho & \pi \\ \xi & A
\end{pmatrix}
\end{equation}
where 
\begin{equation}
\rho = x_1x_2x_3\Bigl( \frac{3 (\gamma+\beta) -\alpha\psi}{\alpha^2}\Bigr)
\end{equation}
and
\begin{equation}
\pi = \xi^t = \Bigl(x_1^2 + \frac{\beta+\gamma}\alpha x_2x_3,
x_2^2+\frac{\beta+\gamma}\alpha x_1x_3,
x_3^2 +\frac{\beta+\gamma}\alpha x_1x_2
\Bigr)
\end{equation}
Again, the limit $\alpha\to 0$ is smooth in view of \eqref{limit}, and
is of the form
\begin{equation}
\tilde f \sim
\begin{pmatrix}
0 & x_1^2 & x_2^2 & x_3^2 \\
x_1^2 & 0 & x_3 &  -x_2 \\
x_2^2 & -x_3 & 0 & x_1 \\
x_3^2 & x_2 & -x_1 & 0 
\end{pmatrix}
+ \alpha
\begin{pmatrix}
-x_1x_2x_3 & 0 & 0 & 0\\
0 & x_1 & 0 & 0\\
0 & 0 & x_2 &0\\
0 & 0 & 0  & x_3 
\end{pmatrix}
+\calo(\alpha^2)
\end{equation}

It is easy to see that the resulting $Q=\left(\begin{smallmatrix} 0 & \tilde f\\
\tilde g &0\end{smallmatrix}\right)$ is equivalent (up to CP signs) with the
tensor product of minimal model branes $M_1(x_i)^{\otimes i=1,2,3}$
deformed by $\Phi$.

\subsection{Moduli space of branes}
\label{tormoduli}

We have seen that for any $(\alpha,\beta,\gamma)$ obeying
$$
\alpha^3+\beta^3+\gamma^3 +\psi \alpha\beta\gamma = 0,
$$
we have a $4\times 4$ matrix factorization of $W$ in (\ref{torus}).
The scaling
$(\alpha,\beta,\gamma)\to (u\alpha,u\beta,u\gamma)$ corresponds to
a basis change of Chan-Paton factors. For example,
in the region $\alpha\beta\gamma\ne 0$, the scaling corresponds to
$$
f\to \begin{pmatrix}
1&\\
&u{\bf 1}_3
\end{pmatrix}f
\begin{pmatrix}
u^{-1}&\\
&{\bf 1}_3
\end{pmatrix},
\quad
g\to
\begin{pmatrix}
u&\\
&{\bf 1}_3
\end{pmatrix}g
\begin{pmatrix}
1&\\
&u^{-1}{\bf 1}_3
\end{pmatrix},
$$
which is done by a certain scaling of the basis elements of
the Chan-Paton factor.
It is a simple excercise to show this also
in the regions near $\alpha\beta\gamma=0$.
Thus, matrix factorizations for a fixed $\psi$ are parameterized 
by the torus of modulus $\psi$ itself.
Namely, the moduli space of the branes
for a given torus is the torus itself!

Note that we need to make a basis change (\ref{basischange})
as we approach the point $[\alpha,\beta,\gamma]=[0,1,-1]$,
and similarly near the eight other points with
$\alpha\beta\gamma=0$.
This suggests that the supercharge $Q$ is not a holomorphic function
on the moduli space, but is a section of a 
certain bundle.
In the context of string field theory, supercharge $Q$ defines
the BRST operator. This is an interesting circumstance where
non-trivial gauge transformations of the BRST operator play an 
important role.
We expect that this property holds for more general Gepner models including
those with $\hat{c}=3$ that are relevant for ${\mathcal N}=1$
compactifications.

\subsection{Geometric interpretation}

As we have seen, the above family of branes are finite
deformations of the brane
$M_1(x_1)\otimes M_1(x_2)\otimes M_1(x_3)$ at $\psi=0$,
which is identified as the ${\bf L}=(0,0,0)$
rational brane in CFT.\footnote{We have been ignoring the label
$m=1,3,5$ that specifies the $\zet_3$ action on the Chan-Paton factor,
but the above story holds in all values of $m$.
They correspond to ${\bf L}=(0,0,0)$ and $M=m-1$ branes.}
The geometrical interpretation of such rational branes are studied in
\cite{BDLR} by computing the charges. According to their results,
it can be interpreted as a brane wrapped
on the torus itself and supporting a holomorphic
bundle of trivial topology.
The brane is specified by the holomorphic structure of the line bundle.
Thus, the moduli space is the Jacobian of the torus, which is the same
as the torus itself as a complex manifold.
We have seen that this is indeed the case for the Landau-Ginzburg branes.

In \cite{laza}, a direct way to obtain more precise geometry of the brane
is described. The prescription of \cite{laza} is first to view a matrix
factorization, as they were originally introduced in \cite{eisen}, as
defining a free resolution of a so-called maximal Cohen-Macaulay module
$\calm$ over the ring $R = \complex[x_i]/W$,
\begin{equation}
\cdots \overset{B}\longrightarrow R^3 \overset{A}\longrightarrow R^3 
\overset{B}\longrightarrow R^3 
\overset{A}\longrightarrow R^3 
\longrightarrow\calm \longrightarrow 0\,,
\eqlabel{longexact}
\end{equation}
\ie, $\calm={\rm Coker} A$. The second step of \cite{laza} is to 
consider the sheafification of $\calm$, after which it obtains a 
geometric interpretation as a bundle over the elliptic curve defined by
the vanishing of the polynomial \eqref{torus} in $\projective^2$.
This procedure is very reminiscent of the gauged linear sigma model 
philosophy. It appears to make sense in more general Gepner models as
well, and it would be very important to verify it from the physics 
point of view.

\section{The quintic}

In this section, we consider the LG model based on the general 
quintic superpotential
\begin{equation}
W = x_1^5+x_2^5+x_3^5+x_4^5+x_5^5 + {\rm deformations}
\eqlabel{quintic}
\end{equation}
and orbifolded by a diagonal $\zet_5$. When all deformations 
vanish, we can obtain matrix factorizations by taking tensor products 
of minimal model branes. We are interested in studying the behavior
of these factorizations under open and closed string deformations.

\subsection{A simple example}

Let us consider
\begin{equation}
M = M_1(x_1)\otimes M_1(x_2)\otimes M_1(x_3)\otimes M_1(x_4)
\otimes M_1(x_5)
\eqlabel{brane}
\end{equation}
where the tensor product is taken in the graded sense. ($M$ 
corresponds to the ${\bf L}=(0,0,0,0,0)$ Recknagel-Schomerus
\cite{RS} brane in CFT.) After orbifolding, the open string spectrum of $M$
consists of one bosonic operator (the identity), and
one fermionic, which is given as a tensor product
\begin{equation}
\Phi=\phi^1_{1,1,j=0}(x_i)^{\otimes i=1,\ldots 5}
\end{equation}
In distinction to the case of the torus, $\Phi$, which is the 
"Serre dual" of the identity, has $U(1)$ charge $3$, and does not
lead to a marginal deformation. So, $M$ is rigid, and does
not have any moduli space.

Instead, let us ask the question what happens to $M$ under
deformations of the bulk. Consider adding a degree $5$ monomial
to $W$,
\begin{equation}
W(\psi)= x_1^5+x_2^5+x_3^5+x_4^5+x_5^5 + \psi \prod x_i^{m_i}
\end{equation}
with $\sum m_i=5$. To be specific, we take $x_1x_2x_3x_4x_5$,
but other cases work similarly. We are looking for a deformation
\begin{equation}
Q = Q_0 +  \Delta Q
\end{equation}
where $Q_0$ is obtained from taking the tensor product of 
$M_1$ in five minimal models.
\begin{equation}
Q_0 = Q_1(x_1)+Q_1(x_2) + Q_1(x_3)+Q_1(x_4)+Q_1(x_5) \,,
\end{equation}
where 
\begin{equation}
Q_1(x) = \begin{pmatrix} 0 & x\\ x^4 & 0\end{pmatrix} \,,
\end{equation}
and it is understood that $Q_1(x_i)$ only operates in the
$i$-th factor in \eqref{brane}.
Imposing $Q^2=W(\psi)$ leads to the equation
\begin{equation}
\{Q_0,Q_1\} + (\Delta Q)^2 = \psi x_1x_2x_3x_4x_5
\end{equation}
A solution to this equation is
\begin{equation}
\Delta Q = \psi 
\begin{pmatrix} 0 & 0 \\ 1 & 0\end{pmatrix}
\otimes x_2 {\rm id} \otimes \cdots \otimes x_5 {\rm id}
\end{equation}
Indeed
\begin{equation}
\biggl\{Q_1(x_1),
\begin{pmatrix} 0 & 0\\1&0\end{pmatrix}
\biggr\} = 
\biggl\{
\begin{pmatrix}
0 & x_1\\x_1^4 & 0
\end{pmatrix},
\begin{pmatrix} 0 & 0\\1&0\end{pmatrix}
\biggr\}
= x_1\;{\rm id}
\end{equation}
\begin{equation}
\bigl\{ Q_1(x_2), \Delta Q\bigr\} = 0
\end{equation}
etc., and
\begin{equation}
\begin{pmatrix}
0 & 0\\1&0
\end{pmatrix}^2
=0
\end{equation}

Of course, we could have chosen a different solution
\begin{equation}
\widetilde{\Delta Q} =
x_1{\rm id} \otimes 
\begin{pmatrix} 0 & 0\\1 & 0\end{pmatrix}
\otimes \cdots \otimes x_5 {\rm id}
\end{equation}
but $\Delta Q$ and $\widetilde{\Delta Q}$ are simply related by a
gauge transformation.
\begin{equation}
\Delta Q - \widetilde{\Delta Q} = 
\biggl\{ Q_0,
\begin{pmatrix}
0 & 0 \\1 & 0
\end{pmatrix}
\otimes
\begin{pmatrix}
0 & 0 \\1&0
\end{pmatrix}
\otimes
x_3{\rm id}\otimes\cdots\otimes x_5{\rm id}
\biggr\}
\end{equation}

As mentioned above, an analogous deformation of the boundary exists for 
deformation of the bulk by any monomial of degree $5$. Moreover, these
deformations mutually (anti-)commute, so we obtain a factorization of
a generic quintic superpotential which reduces for the Fermat
quintic to the tensor prouct of minimal model branes.

\subsection{An obstructed deformation}

Consider the brane 
\begin{equation}
M = M_2(x_1)\otimes M_2(x_2)\otimes M_2(x_3)\otimes M_2(x_4)\otimes M_2(x_5)
\end{equation}
which corresponds to the ${\bf L}=(1,1,1,1,1)$ boundary state in CFT.
In a single minimal model with $k=3$, the brane $M_2$ has
two bosonic and two fermionic boundary fields with the following
charges
\begin{align}
\phi^0_0 &=\begin{pmatrix} 1&0\\0&1     \end{pmatrix} & q=0\\
\phi^1_1 &=\begin{pmatrix} 0&1\\ -x&0   \end{pmatrix} & q=\frac15\\
\phi^0_1 &=\begin{pmatrix} x&0\\0&x     \end{pmatrix} &q=\frac25\\
\phi^1_0 &=\begin{pmatrix} 0&x\\ -x^2&0 \end{pmatrix}&q=\frac35
\eqlabel{spectrum}
\end{align}
Note that the spectrum of charges is identical to the bulk, where we have 
the $(c,c)$ ring $(0,x,x^2,x^3)$ with the charges $q_L=q_R=(0,\frac15,
\frac25,\frac35)$, respectively. 

Therefore, when we take tensor
product and orbifold, the projections work in the same way in the 
bulk and on the boundary. In the ordinary $\zet_5$ orbifold, 
the brane $M$ has $101$ fermionic operators with charge $q=1$, and
$101$ bosonic operators with charge $q=2$ (and in addition
one operator of charge $0$ and one of charge $3$, but they are 
not important here). In the $(\zet_5)^4$ orbifold, which is
the mirror quintic, $M$ has one fermionic operator
\begin{equation}
\Phi = \phi^1_1(x_i)^{\otimes i=1,\ldots, 5}
\end{equation}
and one bosonic operator
\begin{equation}
\Phi^\dagger = \phi^0_1(x_i)^{\otimes i=1,\ldots, 5} = x_1x_2x_3x_4x_5 \, {\rm id}^{\otimes 5}
\end{equation}
We note that $\Phi^\dagger$ is proportional to the identity matrix. It can
be viewed either as a boundary field, where it corresponds to a
charge $q=2$ field in the cohomology of $M$, or as a bulk field,
where it is an element of the $(c,c)$ ring of left-right
charge $(1,1)$. Let us distinguish the bulk field by denoting it as
$\Psi$. We note that
\begin{equation}
\Phi^2 = -\Phi^\dagger = - \Psi\,{\rm id} \,.
\end{equation}
Consider deforming $Q$ by $\Phi$
\begin{equation}
Q \to Q(\varphi) = Q + \varphi \Phi \,.
\eqlabel{modi}
\end{equation}
$Q(\varphi)$ fails to square to $W$ at order $\varphi^2$,
\begin{equation}
\bigl(Q(\varphi)\bigr)^2 - W= - \varphi^2 \Phi^\dagger
\end{equation}
and we can ask whether we can improve $Q(\varphi)$ at order
$\varphi^2$ in order to fix this. However, this
would require a field $\tilde\Phi$ with the property
\begin{equation}
\{Q,\tilde\Phi\} = -\Phi^2 = \Phi^\dagger
\eqlabel{obstruct}
\end{equation}
But such a field cannot exist, because $\Phi^\dagger$ is non-trivial 
in the cohomology of $Q$. We conclude that the deformation
is obstructed at order $\varphi^2$.

On the other hand, if we consider deforming the bulk
\begin{equation}
W \to W(\psi)=W+\psi\Psi \,,
\end{equation}
we can modify $Q$ by $\Phi$ as in \eqref{modi}.
Imposing $Q(\varphi)^2 = W(\psi)$ leads to the equation
\begin{equation}
\varphi^2 +\psi=0.
\label{Fmirq}
\end{equation}
This is the F-term equation of the system.
For each $\psi\ne 0$, it has two solutions for
$\varphi$, and we note immediately
that the two solutions are not gauge equivalent because
$\Phi$ is not exact. 

\subsection{Superpotential and a mirror interpretation}

To give a geometric interpretation to \eqref{Fmirq}, we
need to recall the geometric objects that correspond to
the LG branes we have been studying. The Gepner to large 
volume mapping of the charge lattice for B-type branes
on the quintic was discussed in \cite{BDLR}. One of the 
$5$ branes with ${\bf n}=(2,2,2,2,2)$ in the $\zet_5$ orbifold 
of \eqref{quintic} corresponds at large volume to a rank $8$ 
bundle with Chern character $8-4H - 4H^2+\frac 73H^3$, where $H$ 
is the hyperplane of $\projective^4$. The interest of this 
brane is that it is the anomaly cancelling bundle for 
Type I string theory on quintic \cite{orientifold}
(with non-trivial action on the B-field). It would be 
interesting to check geometrically the fact that this
bundle has $101$ first order deformations, and that the
one associated with $x_1x_2x_3x_4x_5$ is obstructed as in
\eqref{Fmirq}. 

A different interpretation results by considering the factorization
$M$ in the mirror quintic, the $(\zet_5)^4$ orbifold of 
\eqref{quintic}. There are $625$ branes of this type, which
via mirror symmetry correspond precisely to $625$ 
rational A-type RS branes in the quintic model. As also
shown in \cite{BDLR,orientifold}, these $625$ branes are 
identified at large volume with the $625$ special Lagrangian 
submanifolds obtained as real quintics. Our F-term constraint 
\eqref{Fmirq} provides a definite solution of a certain puzzle that 
has accompanied this identification. Topologically, a real
quintic is nothing but an $\reals\projective^3$, and as such
is geometrically rigid. Nonetheless, there are two choices of flat 
$U(1)$ bundle coming from $\pi_1(\reals\projective^3)=\zet_2$ 
\cite{BDLR}. These two supersymmetric branes for fixed K\"ahler 
class (which corresponds to $\psi$ in this context) precisely 
match with the two solutions of (\ref{Fmirq}). The fact that 
the brane on $\reals\projective^3$ develops a massless field 
$\Phi$ as it is continued to small volume is something that is 
not predictible using classical geometry, and it would be 
interesting to understand how it comes about.

Let us make some comments on the spacetime superpotential ${\mathcal W}$.
It must be a holomorphic function of $(\psi,\varphi)$
such that the solution to
${\mathcal W}=\partial_{\psi}{\mathcal W}=\partial_{\varphi}{\mathcal W}=0$
is given by (\ref{Fmirq}). It must also be invariant under the
$\zet_5$ identification,
$(\psi,\varphi)\to (\alpha \psi,\alpha^{-2}\varphi)$, $\alpha^5=1$,
that is induced from the change of variables $(x_1,x_2,x_3,x_4,x_5)
\to (\alpha x_1,x_2,x_3,x_4,x_5)$. One satisfying these
constraints is
${\mathcal W}=f((\psi+\varphi^2)^5)$,
where $f(x)$ is a function such that
$f(x)=f'(x)=0$ has no solution except possibly $x=0$.
Note that this form of the superpotential is consistent with 
the fact that both $\psi$ and $\varphi$ are massless at
$\psi=\varphi=0$.
There is another, global constraint ---
the superpotential is a section of the line bundle ${\mathcal L}$
of the moduli space determined from the K\"ahler class \cite{WB}.
Thus, it is important to know the global structure of the moduli space
as well as the complex structure of the line bundle
${\mathcal L}$. It would be interesting if the non-trivial global structure
of the supercharge $Q$ on the moduli space,
as the one we have seen in the torus example (Section~\ref{tormoduli}),
has something to do with
the line bundle ${\mathcal L}$. We hope to clarify this point in a future
work.

\section{Other models}

In this final section, we illustrate in two further examples
some other general features discussed in section \ref{general}.
We will see that the obstruction can generically appear
at higher order in the perturbation, that there can be 
unobstructed boundary deformations also in the case 
$\hat c =3$, as well as the fact that the presence
of D-branes can obstruct marginal bulk deformations.

\subsection{Obstruction of bulk deformations by D-branes}
\label{Obulk}

Consider the ``two-parameter'' model $\projective_{1,1,2,2,2}[8]$.
The LG potential is
\begin{equation}
W = x_1^8 + x_2^8 + x_3^4 + x_4^4 + x_5^4 + {\rm deformations}
\eqlabel{twopar}
\end{equation}
For simplicity, we will consider the orbifold of \eqref{twopar}
by the maximal Greene-Plesser orbifold group, $\Gamma=\zet_8\times
\zet_8\times\zet_4\times\zet_4$. This leaves only two
marginal bulk deformations
\begin{equation}
\begin{split}
\Psi_{1} &= x_1 x_2 x_3 x_4 x_5 \\
\Psi_{2} &= x_1^4 x_2^4 \,.
\end{split}
\end{equation}
The branes associated with the factorization (for simplicity,
we again omit the labels for the orbifold group action),
\begin{equation}
M = M_2(x_1)\otimes M_2(x_2)\otimes M_2(x_3)
\otimes M_2(x_4)\otimes M_2(x_5)
\end{equation}
shows some interesting properties. There is only one marginal
operator invariant under $\Gamma$,
\begin{equation}
\begin{split}
\Phi &= \phi^1_1(x_i)^{\otimes i=1,\ldots,5} \\
&=
\begin{pmatrix} 0 & 1 \\ -x_1^4 & 0 \end{pmatrix}
\otimes
\begin{pmatrix} 0 & 1 \\ -x_2^4 & 0 \end{pmatrix}
\otimes
\begin{pmatrix} 0 & 1 \\ -1 & 0 \end{pmatrix}
\otimes
\begin{pmatrix} 0 & 1 \\ -1 & 0 \end{pmatrix}
\otimes
\begin{pmatrix} 0 & 1 \\ -1 & 0 \end{pmatrix}
\end{split}
\end{equation}
with conjugate obstruction
\begin{equation}
\Phi^\dagger = \Psi_1 \cdot {\rm id}
\end{equation}
and
\begin{equation}
\Phi^2 = - \Psi_2 \cdot {\rm id} \,.
\end{equation}

Thus, $\Phi^2$ as well as the marginal bulk deformation
$\Psi_2$ are exact on the boundary. The deformation problem
therefore has a solution up to second order. When 
constructing the solution to higher order, as in 
\eqref{higher}, one soon notices that $Q_n$ is only 
non-trivial in the first two minimal model factors, 
whereas in the last three factors (those with $h_i=4$), it is either 
proportional to the identity or to $\phi^1_1$, which are 
both independent of $x_3,x_4,x_5$. Thus, the obstruction 
$\Phi^\dagger=\Psi_1$ can actually never appear in the 
perturbative series, and we conclude that the first 
order deformation by $\Phi$ is unobstructed. One can 
also construct the finite deformation explicitly (for abitrary 
$\psi_2$), much as in the case of the torus.

The natural question that then arises is ``What does the
obstruction $\Phi^\dagger$ actually obstruct?'' We claim
that it actually expresses the fact that, in the presence 
of $M$, $\Psi_1$ is not an allowed bulk deformation anymore.
To justify the claim, we have to show that the equation
\begin{equation}
\{ Q,A\} + A^2 = \psi_1\Psi_1\,{\rm id} = \psi_1 x_1x_2x_3x_4x_5\,{\rm id}
\eqlabel{impossible}
\end{equation}
where $Q$ is the supercharge corresponding to $M$ and
$A$ is an arbitrary fermionic field with charge $1$,
has no solution when $\psi_1\neq 0$.

Strictly speaking, we should emphasize that showing that 
\eqref{impossible} has no solution does not exclude the 
possibility that we can find a solution of $P^2=W$ for non-zero 
$\psi_1$ which for $\psi_1\to 0$ does not reduce to the tensor 
product solution $Q$, but is still continuously connected to
it via massive deformations. In other words, the moduli 
space of our brane could have several branches at $\psi_1=0$, 
around some of which $\Psi_1$ is a valid deformation. 
We will here only consider the problem around the rational 
point.

To show that $x_1x_2x_3x_4x_5\cdot {\rm id}$ is not 
contained in the LHS of \eqref{impossible}, we
expand $A$ as
\begin{equation}
A = \sum_j \alpha_j A^j \,,
\eqlabel{Aexp}
\end{equation}
where the $A^j$ form a basis of $\Gamma$-equivariant 
fermionic maps in $\Hom(M,M)$ with homogeneous degree 
$q=1$. Note that even without going to the kernel of $Q$, 
this space is finite dimensional. 

Since $Q$ is at least quadratic in the $x_i$'s, the 
first term in \eqref{impossible} cannot contain $\Psi_1$.
To check the second term, we notice that we can focus
on those $A^j$ in \eqref{Aexp} which are at most
linear in the $x_i$'s. Imposing in addition invariance
under $\Gamma$ and a total R-charge of $1$ in fact
leaves only very few possibilities. Basis elements
$A^j$ must be of one of the forms
\begin{equation}
A^1 =
\begin{pmatrix} 0 & 1\\ 0 & 0 \end{pmatrix} \otimes
\begin{pmatrix} 0 & 1\\ 0 & 0 \end{pmatrix} \otimes
\begin{pmatrix} 0 & a_3 \\ b_3 & 0  \end{pmatrix} \otimes
\begin{pmatrix} 0 & a_4 \\ b_4 & 0  \end{pmatrix} \otimes
\begin{pmatrix} 0 & a_5 \\ b_5 & 0  \end{pmatrix} 
\end{equation}
or
\begin{equation}
A^2 =
\begin{pmatrix} 0 & 0 \\ x_1 & 0\end{pmatrix} \otimes
\begin{pmatrix} 0 & 0 \\ x_2 & 0\end{pmatrix} \otimes
\begin{pmatrix} 0 & c_3 x_3 \\ d_3 x_3 & 0 \end{pmatrix} \otimes
\begin{pmatrix} 0 & c_4 x_4 \\ d_4 x_4 & 0 \end{pmatrix} \otimes
\begin{pmatrix} 0 & c_5 x_5 \\ d_5 x_5 & 0 \end{pmatrix}
\end{equation}
where $a_i$, $b_i$, $c_i$ and $d_i$ are abitrary scalars.
In the second term on the LHS of \eqref{impossible}, this
results in expressions of the form
\begin{equation}
\{ A^1,A^2\} = x_1x_2x_3x_4x_5
\left[
\begin{pmatrix} 1 & 0 \\ 0 & 0
\end{pmatrix} \otimes
\begin{pmatrix} 1 & 0 \\ 0 & 0
\end{pmatrix} +
\begin{pmatrix} 0 & 0 \\ 0 & 1
\end{pmatrix} \otimes
\begin{pmatrix} 0 & 0 \\ 0 & 1
\end{pmatrix}
\right] \otimes \cdots \,,
\end{equation}
which clearly falls short of $\Psi_1\cdot{\rm id}$.

To summarize, the sole F-term constraint in the present model is
\begin{equation}
\psi_1=0.
\end{equation}

\subsection{Obstruction at higher order}

Finally, we consider a slightly more complicated model which
combines several of the previous features.

The model under consideration is a popular ``three-parameter'' 
model with
\begin{equation}
W = x_1^{15} + x_2^{5} + x_3^5 + x_4^5 + x_5^3 + {\rm deformations}
\end{equation}
and where we again orbifold by the maximal group $\zet_{5}
\times \zet_5\times\zet_5\times\zet_3$. Marginal bulk deformations
are
\begin{equation}
\begin{split}
\Psi_1 &= x_1x_2x_3x_4x_5 \\
\Psi_2 & = x_1^{10} x_5 \\
\Psi_3 &= x_1^6 x_2x_3x_4
\end{split}
\end{equation}
The tensor product brane with ${\bf n}=(7,2,2,2,1)$ has three 
invariant marginal boundary deformations
\begin{equation}
\begin{split}
\Phi_1 &=
\begin{pmatrix} 0 & 1\\-x_1 & 0\end{pmatrix} \otimes
\begin{pmatrix} 0 & 1\\-x_2 & 0\end{pmatrix} \otimes
\begin{pmatrix} 0 & 1\\-x_3 & 0\end{pmatrix} \otimes
\begin{pmatrix} 0 & 1\\-x_4 & 0\end{pmatrix} \otimes
\begin{pmatrix} 0 & 1\\-x_5 & 0\end{pmatrix} \\
\Phi_2 &=
x_1^5{\rm id} \otimes {\rm id} \otimes {\rm id} \otimes {\rm id}
\otimes 
\begin{pmatrix} 0 & 1\\-x_5 & 0\end{pmatrix} \\
\Phi_3 &=
x_1^3{\rm id} \otimes
\begin{pmatrix} 0 & 1\\-x_2 & 0\end{pmatrix} \otimes
\begin{pmatrix} 0 & 1\\-x_3 & 0\end{pmatrix} \otimes
\begin{pmatrix} 0 & 1\\-x_4 & 0\end{pmatrix} \otimes
{\rm id}
\end{split}
\end{equation}
The obstructions are
\begin{equation}
\begin{split}
\Phi_1^\dagger &= x_1^6 x_2x_3x_4\, {\rm id} \\
\Phi_2^\dagger &= 
\begin{pmatrix} 0 & x_1\\-x_1^2 & 0\end{pmatrix} \otimes
\begin{pmatrix} 0 & x_2\\-x_2^2 & 0\end{pmatrix} \otimes
\begin{pmatrix} 0 & x_3\\-x_3^2 & 0\end{pmatrix} \otimes
\begin{pmatrix} 0 & x_4\\-x_4^2 & 0\end{pmatrix} \otimes {\rm id}\\
\Phi_3^\dagger &=
\begin{pmatrix} 0 & x_1^3\\-x_1^4 & 0\end{pmatrix} \otimes
x_2{\rm id}\otimes x_3{\rm id}\otimes x_4{\rm id} \otimes
\begin{pmatrix} 0 & 1\\-x_5 & 0\end{pmatrix} 
\end{split}
\end{equation}
Also,
\begin{equation}
\begin{split}
\Phi_1^2 &= -\Psi_1{\rm id} \\
\Phi_2^2 &= -\Psi_2{\rm id} \\
\Phi_3^2 &= -\Psi_3{\rm id} = -\Phi_1^\dagger
\end{split}
\end{equation}
Thus $\Phi_3$ behaves much like the obstructed deformation 
we studied in the quintic case. $\Phi_2^2$ or $\Psi_2$ are 
exact on the boundary, and the obstructions all involve
$x_2$, $x_3$, $x_4$ non-trivially, so this case is similar to
the one in the previous subsection, and $\Phi_2$ is not obstructed.
Consider, finally, deformations by $Q_1=\Phi_1$. To second order,
we can write,
\begin{equation}
\Phi_1^2 = \Bigl\{ Q, \underbrace{- x_1{\rm id}
\otimes x_2{\rm id}\otimes x_3{\rm id}\otimes x_4{\rm id}\otimes
\begin{pmatrix} 0&0\\1&0\end{pmatrix}}_{=:Q_2} \Bigr\}
\end{equation}
We then find
\begin{equation}
\{Q_2,Q_1\} = \Phi_2^\dagger
\end{equation}
so that here the obstruction appears at third order in the deformation.
One may also note that the three first order deformation
$\Phi_1$, $\Phi_2$ and $\Phi_3$ do not mutually commute. As a
consequence, their joint deformation problem is more
intricate.

\begin{acknowledgments} We would like to thank Ragnar Buchweitz,
Michael Haack, Anton Kapustin, Calin Lazaroiu, Wolfgang Lerche,
Erich Poppitz, Radu Roiban, Ed Witten for instructive discussions.
The work of KH was supported by Alfred P. Sloan Foundation,
Connaught Foundation
and NSERC. The research of JW was supported in part by the National Science 
Foundation under Grant No.\ PHY99-07949.

\end{acknowledgments}



\providecommand{\href}[2]{#2}\begingroup\raggedright

\end{document}